\documentclass{article}

\usepackage{amsmath, amsthm, amssymb}
\usepackage{graphicx}
\usepackage{verbatim}
\usepackage{natbib}
\usepackage{caption}
\usepackage{subcaption}
\usepackage{fancyvrb}
\usepackage{enumerate}
\usepackage{enumitem}
\usepackage{relsize}
\usepackage{hyperref}
\usepackage[margin=1.5in]{geometry}
\hypersetup{colorlinks,citecolor=blue,urlcolor=blue,linkcolor=blue}
\usepackage{diagbox}
\usepackage{stefan_tex}
\usepackage{float}

\usepackage{multirow}

\usepackage[utf8]{inputenc}
\usepackage[english]{babel}

\usepackage{accents}

\usepackage{pgfplots}
\usepackage{tikzscale}
\pgfplotsset{compat=newest}
\usepgfplotslibrary{groupplots}
\usepgfplotslibrary{polar}
\usepgfplotslibrary{smithchart}
\usepgfplotslibrary{statistics}
\usepgfplotslibrary{dateplot}
\usepgfplotslibrary{external}
\usetikzlibrary{arrows.meta}
\usetikzlibrary{backgrounds}
\usepgfplotslibrary{patchplots}
\usepgfplotslibrary{fillbetween}
\pgfplotsset{%
layers/standard/.define layer set={%
    background,axis background,axis grid,axis ticks,axis lines,axis tick labels,pre main,main,axis descriptions,axis foreground%
}{grid style= {/pgfplots/on layer=axis grid},%
    tick style= {/pgfplots/on layer=axis ticks},%
    axis line style= {/pgfplots/on layer=axis lines},%
    label style= {/pgfplots/on layer=axis descriptions},%
    legend style= {/pgfplots/on layer=axis descriptions},%
    title style= {/pgfplots/on layer=axis descriptions},%
    colorbar style= {/pgfplots/on layer=axis descriptions},%
    ticklabel style= {/pgfplots/on layer=axis tick labels},%
    axis background@ style={/pgfplots/on layer=axis background},%
    3d box foreground style={/pgfplots/on layer=axis foreground},%
    },
}

\usepackage{algorithm, algorithmic}

\usepackage{soul}
\usepackage{dsfont}

\graphicspath{{./figures/}}

\theoremstyle{plain}
\newtheorem{prop}{Proposition}

\newtheorem{coro}[prop]{Corollary}

\theoremstyle{definition}

\theoremstyle{remark}

\newcommand{\Var}{\text{Var}}
\newcommand{\E}{\mathbb{E}}
\newcommand{\R}{\mathds{R}}
\renewcommand{\P}{\mathbb{P}}

\author{
\makebox[45mm]{Kevin Wu Han} \\  Stanford University \and
\makebox[45mm]{Han Wu} \\ Stanford University}

\date{Draft version \ifcase\month\or
January\or February\or March\or April\or May\or June\or
July\or August\or September\or October\or November\or December\fi \ \number%
\year\ \  }

\title{Ensemble Method for Estimating Individualized Treatment Effects \thanks{Authors
are listed in alphabetical order. We thank
Kevin Guo
for helpful discussions.}}

\begin{document}

\maketitle

\begin{abstract}
In many medical and business applications, researchers are interested in estimating individualized treatment effects using data from a randomized experiment. For example in medical applications, doctors learn the treatment effects from clinical trials and in technology companies, researchers learn them from A/B testing experiments. Although dozens of machine learning models have been proposed for this task, it is challenging to determine which model will be best for the problem at hand because ground-truth treatment effects are unobservable.  In contrast to several recent papers proposing methods to \textit{select} one of these competing models, we propose an algorithm for \textit{aggregating} the estimates from a diverse library of models.  We compare ensembling to model selection on 43 benchmark datasets, and find that ensembling wins almost every time.  Theoretically, we prove that our ensemble model is (asymptotically) at least as accurate as the best model under consideration, even if the number of candidate models is allowed to grow with the sample size.
\end{abstract}

\section{Introduction}\label{sec:intro}
The conditional average treatment effect (CATE) function $\tau (\cdot)$ is an object of central importance in precision medicine and targeted marketing \citep{uplift_review, personalized_medicine}.  For any feature profile $x$, $\tau(x)$ gives the average effect of a treatment (relative to control) among individuals with features $x$.  Estimates of $\tau(\cdot)$ can be used to target a drug or advertisement to the subpopulation where it will have the largest impact.

In recent years, causal inference researchers have proposed dozens of methods that adapt machine learning algorithms so that they can be used to estimate CATE functions from experimental data.  For example, random forests, gradient boosting, and deep neural networks have all been repurposed for this task \citep{causal_forests, causal_boosting, deep_learning_ite, shalit_johansson_sontag}.  Going further, several papers have proposed general-purpose ``meta-learning frameworks'' that allow analysts to adapt \textit{any} supervised learning algorithm to estimate CATE functions \citep{meta_learners, quasi_oracle}.  Today, fitting half a dozen different CATE models is only a few hours of work for a skilled data analyst.

That being said, deciding \textit{which} of these competing models to use on the problem at hand remains challenging.  Evidence from large-scale simulation studies and causal inference competitions show that no single method performs the best in every problem \citep{meta_learners, comparison, causal_tournament}.  Unfortunately, standard model selection techniques from supervised learning cannot be applied to solve this problem, since individualized treatment effects are unobservable.

To address this problem, several authors have proposed ``causal'' variants of cross-validation which can be used for model selection \citep{influence_functions, synth_validation}.  In contrast, we propose a method that bypasses model selection entirely and instead computes an ensemble of candidate models. For bounded outcomes, we show that no validation-set-based model selection rule can (asymptotically) outperform our ensemble model.  This result holds even if the number of candidate models grows faster than the sample size, and we do not make any assumptions about the quality or correlation structure of the candidate models.  We assess the empirical performance of ensembling by applying our method to 43 benchmark semi-synthetic datasets, and find that ensembling almost always outperforms model selection.  Our practical recommendation for data analysts estimating CATE models is to fit a wide variety of diverse models, and then use ensembling rather than model selection to obtain a CATE model for deployment.

\section{Formal Problem Statement}

\subsection{The potential outcomes framework}
We adopt potential outcomes model of causality introduced by Neyman and Rubin \citep{neyman_rubin}.  In this model, we posit that, in addition to an individual's observed features $X_i \in \mathcal{X}$, he or she is endowed with two unobserved \textit{potential outcomes} $Y_{1i}, Y_{0i} \in \R$.  Here, $Y_{1i}$ is the outcome we would observe if individual $i$ were assigned to the ``treatment'' condition, and $Y_{0i}$ is the outcome we would observe if individual $i$ were assigned to the ``control'' condition.  The \textit{treatment effect} for individual $i$ is defined as the contrast between these two potential outcomes, $\tau_i = Y_{1i} - Y_{0i}$.  Since only one of the two potential outcomes can be observed for any individual, $\tau_i$ is unobservable.

\subsection{Assumptions on the randomized experiment}
In this paper, we focus on randomized experiments. We do so because of two reasons. First, randomized experiments are ubiquitous in applications. Thousands of A/B tests are being performed in technology companies every day and randomized controlled trials are the gold standard in medical applications. Learning a better model for treatment effects leads to better personalization. Second, estimation from observational data typically relies on much stronger, unverifiable assumptions. It is our goal in this paper to minimize assumptions and keep things simple. 

We assume that the analyst has access to $n$ samples from a randomized experiment that was performed in the following fashion:
\begin{itemize}
    \item[1.] Participants in the experiment were sampled i.i.d. from the population of interest, i.e. $(X_i, Y_{1i}, Y_{0i}) \sim \P$ independently for each $i \in [n]$.
    
    \item[2.] Independently of $\{ (X_i, Y_{1i}, Y_{0i}) \}_{i = 1}^n$, treatment assignments $(Z_1, \cdots, Z_n) \in \{ 0, 1 \}^n$ were sampled from either a Bernoulli design or a Completely Randomized design.  In a Bernoulli design, $\P(Z_i = 1) = p$ independently across $i \in [n]$.  In a completely randomized design, the vector $(Z_1, \cdots, Z_n)$ is sampled uniformly from the set of binary vectors with sum $p n$.  In our notation, $Z_i = 1$ indicates that participant $i$ was assigned to the ``treatment" condition, and $Z_i = 0$ indicates that participant $i$ was assigned to the ``control" condition.
    
    \item[3.] The observed outcome is $Y_i = Z_i Y_{1i} + (1 - Z_i) Y_{0i}$.  The final dataset available to the analyst is $\{ (X_i, Y_i, Z_i) \}_{i = 1}^n$.
\end{itemize}

\subsection{The estimation problem}
We study the problem of estimating the conditional average treatment effect function $\tau(\cdot)$, which can be defined as:
\begin{align}
    \tau(x) := \E[ Y_{1i} - Y_{0i} | X_i = x] = \E[ \tau_i | X_i = x]
\end{align}
Under the experimental assumptions stated above, the function $\tau(\cdot)$ is identified (up to $\P$-almost sure equivalence).

We assume that the analyst has a library of $K$ candidate algorithms for estimating $\tau(\cdot)$, where each algorithm $\mathcal{A}_k$ maps a dataset $\{ (X_i, Y_i, Z_i) \}$ (of any size, not necessarily $n$) to an estimated CATE function $\hat{\tau}_k : \mathcal{X} \rightarrow \R$.  The estimation problem is to use these algorithms $\{ \mathcal{A}_k \}_{k = 1}^K$ to construct a single function $\hat{\tau}_s(\cdot)$ that is close to the true CATE function $\tau(\cdot)$.  We will measure the accuracy of an estimator $\hat{\tau}(\cdot)$ using its $\mathcal{L}^2(\P_X)$ distance to $\tau(\cdot)$, $|| \hat{\tau} - \tau ||_2 = \sqrt{\E[ ( \hat{\tau}(X_{\text{new}}) - \tau(X_{\text{new}}))^2]}$.  This is the most commonly used accuracy measure in the literature on heterogeneous treatment effect estimation, and is called the \textit{precision in estimating heterogeneous effects} by Hill \citep{bayesian_nonparametric}.

\section{Our Proposal}

\subsection{The causal stacking algorithm}

Our proposal, which we call \textit{causal stacking}, can be simply stated.  First, split the data into a training and an averaging set.  Then, use the CATE algorithms $\mathcal{A}_k$ to learn estimated CATE models $\hat{\tau}_k(\cdot)$ on the training set. Finally, use the averaging set to find a weight vector $\hat{w}$ such that $\tau_s = \hat{w}^{\top} \hat{\tau}_{1:K}$ minimizes an estimate of $|| \hat{\tau}_s - \tau ||^2_2$.  A more complete description of the method is given in Algorithm \ref{causal_stacking}\footnote{The non-negativity constraint and $\ell_1$ constraint are for interpretability and help in theoretical analysis. We consider extensions in empirical study section}.

\begin{algorithm}[htb]
   \caption{Causal stacking}
   \label{causal_stacking}
\begin{algorithmic}[1]
   \STATE \textbf{Input:} Data $\{(X_i, Y_i, Z_i)\}_{i = 1}^n$, CATE algorithms $\{ \mathcal{A}_k \}_{k = 1}^K$.
   \STATE Partition the data into a training set $\mathcal{S}_{\text{train}}$ containing $100(1 - \alpha)\%$ of the data and an averaging set $\mathcal{S}_{\text{avg}}$ containing $100 \alpha \%$ of the data.  If the experimental design was Bernoulli, the partition can be random.  If the experimental design was Completely Randomized, the fraction of treated units in the averaging set should be $p$.
   \FOR{$t \in \{ 0, 1 \}$}
   \STATE Using the data in $\mathcal{S}_{\text{train}}$, fit a regression model $\hat{\mu}_t$ that predicts $Y_{ti}$ using $X_i$, i.e. $\hat{\mu}_t$ is an estimate of $\E[Y_i |X_i, W_i = t]$
   \ENDFOR
   \FORALL{$k \in [K]$}
   \STATE Set $\hat{\tau}_k \leftarrow \mathcal{A}_k( \mathcal{S}_{\text{train}})$.
   \ENDFOR
   \STATE Solve the optimization problem
   \begin{align}
   \label{plug_in}
      \hat{w} = \argmin_{w \succeq 0, || w ||_1 = 1} \frac{1}{| \mathcal{S}_{\text{avg}}|} \sum_{i \in \mathcal{S}_{\text{avg}}} ( \hat{\tau}_i - w^{\top} \hat{\tau}_{1:K}(X_i))^2
   \end{align}
   where $\hat{\tau}_i$ is defined by:
   \begin{align}
   \label{aipw}
       \begin{split}
       \hat{\tau}_i &= [ \hat{\mu}_1(X_i) - \hat{\mu}_0(X_i)]\\
       &+ \frac{[Y_i - \hat{\mu}_1(X_i)] Z_i}{p} - \frac{[Y_i - \hat{\mu}_0(X_i)](1 - Z_i)}{1 - p}
       \end{split}
   \end{align}
   \STATE \textbf{Output:} $\hat{\tau}_s(\cdot) = \hat{w}^{\top} \hat{\tau}_{1:K}(\cdot)$.
\end{algorithmic}
\end{algorithm}

\subsection{Practical recommendations}
We have a few practical recommendations regarding the use of Algorithm \ref{causal_stacking}. We will further elaborate in section \ref{simulations}.
\begin{itemize}
    
    \item \textbf{Number of Candidate Models}.  Causal stacking gives more gains when we have larger number of candidate models. For best performance, we recommend trying a wide variety of machine-learning algorithms and meta-learning frameworks.
    
    \item \textbf{Prediction models}.  Although the theoretical guarantees on the causal stacking algorithm do not rely on the correctness or consistency of the models $\hat{\mu}_1$ and $\hat{\mu}_0$, we have found that Algorithm \ref{causal_stacking} performs better when these models have low out-of-sample MSE.  There are two standard approaches to fitting regression models in causal inference:  (i) use only data from treated units to fit $\hat{\mu}_1$ and only data from control units to fit $\hat{\mu}_0$;  (ii) fit a single model $\hat{\mu}(Z_i, X_i)$ using all the data in $\mathcal{S}_{\text{train}}$, then set $\hat{\mu}_1(X_i) = \hat{\mu}(1, X_i)$ and $\hat{\mu}_0(X_i) = \hat{\mu}(0, X_i)$.  We have found that these approaches work well on different datasets, so it is worthwhile to try both and choose the approach with smaller error in cross-validation.
    
    \item \textbf{Refitting on the entire dataset}.  After computing the weight vector $\hat{w}$, one might be tempted to refit the algorithms $\{ \mathcal{A}_k \}$ on the \textit{entire} dataset of $n$ samples to construct estimated CATE models $\{ \hat{\tau}_k^* \}$ which are then averaged using the first-stage weights.  The resulting CATE model $\hat{\tau}_s^*(\cdot) = \hat{w}^{\top} \hat{\tau}^*_{1:K}$ would then be deployed.  Although this approach sounds like it is more sample-efficient, $\hat{\tau}_s^*$ is not guaranteed to perform better than $\hat{\tau}_s$.  CATE models based on trees or neural networks can be quite sensitive to the input dataset and the parameter initialization, so $\mathcal{A}_k( \mathcal{S}_{\text{train}})$ may bear little resemblance to $\mathcal{A}_k( \mathcal{S}_{\text{train}} \cup \mathcal{S}_{\text{avg}})$.
\end{itemize}

\section{Related Work} \label{literature_review}
In supervised learning problems, it is well-known that model averaging could outperform model selection \citep{wolpert1992}.  For example, all of the top submissions in the ``Netflix Prize" competition used extensive ensembling \citep{netflix_prize}, and model averaging is now considered essential to achieve competitive performance in Kaggle tournaments.

Recently, model averaging methods have also made their way into causal inference algorithms, although usually only to optimize a regression model, a propensity score model, or a forecast.  For example, in the panel data context, \citet{athey_ensemble} propose using model ensembling to improve predictions of potential outcomes.  In the causal inference competition of \citet{causal_tournament}, several top-performing submissions used some form of ensembling to optimize potential outcome models, notably the Super Learner + tMLE proposal based on \citet{super_learner} and \citet{tmle}.

The only prior work which we are aware of that proposes stacking for CATE function estimation is Nie \& Wager \citep{quasi_oracle}, where one of the state of art CATE models R-learner is introduced. Their proposal for stacking is based on the decomposition which inspires the R-learner and is quite similar to ours\footnote{We are describing a slightly simplified version of the proposal of Nie \& Wager which uses only a single split of the data.}, but they use the loss function (\ref{R_loss}) in place of the loss function (\ref{plug_in}) and optimize over all nonnegative weight vectors, not just those constrained to sum to one. We call this method R-Stacking. 
\begin{align}
\label{R_loss}
R(w) = \sum_{i \in \mathcal{S}_{\text{avg}}} [Y_i - \hat{\mu}(X_i) - (Z_i - p) (w^{\top} \hat{\tau}_{1:K}(X_i))]^2
\end{align}
In (\ref{R_loss}), $\hat{\mu}$ is any regression function that predicts $Y_i$ using $X_i$.  In balanced experiments $(p = 0.5)$, noting that $(Z_i-p)^2$ is indeed a constant, it is straightforward to show that $R(\cdot)$ is equivalent (up to a constant scaling) to our stacking loss function (\ref{plug_in}) with $\hat{\mu}_1 \equiv \hat{\mu}_0 \equiv \hat{\mu}$. We give a thorough comparison of causal stacking and R-Stacking in Section \ref{simulations}.

The stacking loss function (\ref{plug_in}) in Algorithm \ref{causal_stacking} is simply a feasible approximation of the ``idealized" validation-set estimate of $|| \hat{\tau}_s - \tau ||_2^2$, which would be
\begin{align}
\label{plug_in_cv}
\frac{1}{| \mathcal{S}_{\text{avg}}|} \sum_{i \in \mathcal{S}_{\text{avg}}} ( \tau_i - \hat{\tau}_s(X_i))^2
\end{align}
Since $\tau_i$ is unobservable, we plug in an estimate $\hat{\tau}_i$.  The estimate $\hat{\tau}_i$ that we use was proposed in \citet{aronow_middleton} and shown to be unbiased for $\tau(X_i)$ under any choice of the regression functions $\hat{\mu}_0$ and $\hat{\mu}_1$.  This estimator is closely related to the AIPW/doubly-robust estimator from the literature on observational studies.  The idea of using a plug-in estimator to approximate the idealized validation set procedure has appeared before in the literature on CATE model selection.  For example, \citet{plug_in} and \citet{uplift_modeling} explore this approach with specific choices of $\hat{\mu}_0$ and $\hat{\mu}_1$. \citet{matching_cv} study (\ref{plug_in_cv}) using a matching-based estimator of $\tau_i$.  \citet{influence_functions} propose a perturbation of the plug-in estimate based on influence functions from semiparametric efficiency theory.  On the theoretical side, the analysis in \citet{unified_cv} shows that -- under certain technical conditions -- minimizing the plug-in validation estimate is a consistent model selection rule.

\section{Theoretical Analysis}

In this section, we will state a result saying that -- for bounded outcomes -- replacing the averaging step of Algorithm \ref{causal_stacking} by \textit{any} model selection rule cannot (asymptotically) lead to an improvement, even if the number of candidate models $K$ grows sub-exponentially with the sample size $n$.  The proof is simple, and is based on classical results on the assumption-free risk-consistency of $\ell_1$-constrained functional aggregation \citep{functional_aggregation, presistence}.\\

\begin{prop}
\textbf{(Finite-sample bound)} \label{main_result}\\
Assume that $Y_{0i}, Y_{1i}, \hat{\tau}_k, \hat{\mu}_1$, and $\hat{\mu}_0$ are all uniformly bounded by a constant $L$.  Then, with probability at least $1 - \delta$, the output $\hat{\tau}_s$ of the causal stacking algorithm satisfies:
\begin{align*}
|| \hat{\tau}_s - \tau ||_2^2 & \leq \min_{k \in [K]} || \hat{\tau}_{k} - \tau ||_2^2 + 12 L^2 \sqrt{\frac{\log([K + 1]^2/\delta)}{\alpha n}}
\end{align*}
\end{prop}
\bigskip 
\begin{coro}
\label{asymptotic_implication}
\textbf{(Asymptotic implication)}\\
Assume that $Y_{0i}, Y_{1i}, \hat{\tau}_k, \hat{\mu}_1$ and $\hat{\mu}_0$ are uniformly bounded.  Assume that the number of candidate models $K_n$ grows sub-exponentially with $n$.  Then if $\hat{\tau}_{k^*} \in \{ \hat{\tau}_1, \cdots, \hat{\tau}_K \}$ is the model selected by any model selection rule, we have the asymptotic result:
\begin{align}
\label{asymptotic}
\max \{ 0, || \hat{\tau}_s - \tau ||^2_2 - || \hat{\tau}_{k^*} - \tau ||^2_2 \} \xrightarrow{p} 0
\end{align}
In other words, no model selection algorithm based on a validation set can asymptotically outperform ensembling.
\end{coro}

 The proofs may be found in the supplementary material.  We will make a few remarks about both the assumptions and the conclusions of Proposition \ref{main_result} and Corollary \ref{asymptotic_implication}.
\begin{itemize}
    \item The condition that the regression models $\hat{\mu}_1, \hat{\mu}_0$ and candidate CATE models $\hat{\tau}_k$ are bounded imposes no restriction beyond boundedness of the potential outcomes, since one can always truncate these models to produce a new regression/CATE models which satisfy the boundedness assumption and are at least as accurate as the originals. 
    
    \item The result (\ref{asymptotic}) is most interesting when $\min_k || \hat{\tau}_{k} - \tau ||^2_2$ does \textit{not} tend to zero, i.e. the true CATE function is so complex that no candidate model is consistent.  We doubt that any CATE algorithm can achieve vanishingly small error in any real problem with more than a handful of covariates.  Standard results from statistical minimax theory show that no CATE algorithm $\mathcal{A}_k$ can achieve (worst-case) error better than $\mathcal{O}( n^{-1/2})$ without extremely strong continuity assumptions on $\tau(\cdot)$ unless $\text{dim}( \mathcal{X})$ is very small\footnote{
    Estimating a CATE function is at least as hard as nonparametric regression, since a double $(X_i, Y_i)$ can always be augmented into a triple $(X_i, 0, Y_{1i})$.  The problem of recovering the regression function from the original data is the same as the problem of recovering the CATE function from the augmented data.  Standard results on minimax estimation over nonparametric function classes can then be applied to derive the stated lower bound, c.f. \citet{tsybakov_nonparametric}.}.  This means that, in high-dimensional problems with only moderate smoothness, using $\hat{\tau}_s$ instead of the best candidate model does not degrade the worst-case rate of convergence.
    
    \item Ideally, we would also like to claim that causal stacking is competitive with the best algorithm fit on the \textit{entire} training dataset, whereas Proposition \ref{main_result} only compares it to the best model fit on a $\mathcal{S}_{\text{train}}$.  However, without imposing further assumptions on the algorithms $\mathcal{A}_1, \cdots, \mathcal{A}_K$, it is not possible to prove any such result.  The output of $\mathcal{A}_k$ on a dataset of $n$ samples need not be related in any way to the output of $\mathcal{A}_k$ on a dataset with $\alpha n$ samples.  
\end{itemize}

\section{Empirical Study} \label{simulations}

We performed an extensive simulation study to assess the performance of causal stacking relative to two other approaches:
\begin{enumerate}
    \item \textit{Oracle model selection}.  The oracle model selection procedure selects among candidate models by choosing the model that minimizes the \textit{idealized} validation-set risk estimate (\ref{idealized}).
    \begin{align}
        \label{idealized}
        \frac{1}{| \mathcal{S}_{\text{avg}} |} \sum_{i \in \mathcal{S}_{\text{avg}}} ( \tau(X_i) - \hat{\tau}_k(X_i))^2 
    \end{align}
    In other words, this procedure has oracle knowledge of the true CATE values on the validation set in the model selection step.
    
    \item \textit{R-Stacking}.  The R-Stacking procedure proposed by Nie \& Wager \citep{quasi_oracle} optimizes the loss (\ref{R_loss}) over the class of nonnegative weight vectors.  We use XGBoost model to fit the regression function $\hat{\mu}$ in (\ref{R_loss}).
\end{enumerate}
To facilitate a fair comparison with R-Stacking, we also used XGBoost models to fit the regressions $\hat{\mu}_1$ and $\hat{\mu}_0$ required to implement Algorithm \ref{causal_stacking}.

\subsection{Datasets}

We applied our method to all 43 ``high-heterogeneity" datasets from the 2016 Atlantic Causal Inference Competition \citep{acic2016}.  These semi-synthetic datasets use covariates drawn from a subset of the Collaborative Perinatal Project, but potential outcomes $Y_{1i}, Y_{0i}$ and CATE functions $\tau(\cdot)$ were generated synthetically.  More detail on the construction of these datasets can be found in \citet{causal_tournament}.  We filtered out a small number of unmanageable categorical features, leaving 52 covariates remaining.

For each dataset, we used 2000 observations for training CATE models and 1000 observations for averaging.  The remaining 1802 observations were reserved as a test set for evaluating the averaged models.  In the training and validation data, the assignments $Z_i$ were generated i.i.d. from a Bernoulli($p$) distribution, and we considered $p \in \{ 0.1, 0.3, 0.5 \}$.  Fifty replications were performed per dataset, with both the train/averaging/split and treatment assignments being rerandomized across replications.  Our final quality measure is the average test MSE across the 50 replications.

\subsection{Candidate models}

In our simulations, we considered nine candidate CATE algorithms.  These algorithms span a variety of meta-learning ``frameworks," covering T-, S-, X-, and R-learning \citep{meta_learners, quasi_oracle}.  They also employ a variety of machine learning algorithms.
\begin{enumerate}
    \item \textit{SVM S-learner}.  This method uses the training data to fit a single support vector regression $\hat{\mu}$ making predictions $\hat{Y}_i = \hat{\mu}(Z_i, X_i)$.  We used a Gaussian kernel in the support vector regression.  The fitted CATE function is defined by $\hat{\tau}(x) = \hat{\mu}(1, x) - \hat{\mu}(0, x)$.
    
    \item \textit{XGBoost T-learner}.  This method fits two gradient boosting models $\hat{\mu}_1$ and $\hat{\mu}_0$ using data from treated and control units respectively.  The fitted CATE function is $\hat{\tau}(x) = \hat{\mu}_1(x) - \hat{\mu}_0(x)$.  We used 100 boosting iterations, and all other parameters were set at their default values.
    
    \item \textit{Random Forest T-learner}.  This is the same as the XGBoost T-learner, but the regression functions are fit using random forests.  We used the default parameter settings in the \texttt{ranger} implementation of random forests.
    
    \item \textit{Regression tree S-learner}.  This is the same as the SVM S-learner, but $\hat{\mu}$ is fit using CART.
    
    \item \textit{Lasso T-learner}.  This is the same as the XGBoost T-learner, but the regression functions are fit using Lasso.  The $\ell_1$-penalty parameters were chosen using 5-fold cross-validation.
    
    \item \textit{Random Forest X-learner}.  This method fits two regression models $\hat{\mu}_1$ and $\hat{\mu}_0$ in the same way as the Random Forest T-learner.  After fitting these models, individualized treatment effects $\hat{\tau}_i$ are estimated for each unit by using the regression models to impute the unobserved potential outcomes.  Two new random forest models $\hat{\gamma}_1$ and $\hat{\gamma}_0$ are fit to predict $\hat{\tau}_i$ using $X_i$ based on data from treated and control units, respectively.  The final CATE model is $\hat{\tau}(x) = (1 - p) \hat{\gamma}_1(x) + p \hat{\gamma}_0(x)$.
    
    \item \textit{XGBoost R-learner}.  In this method, we first fit an XGBoost model $\hat{\mu}$ predicting $Y_i$ using $X_i$.  Then, we fit a separate XGBoost model $\hat{\tau}$ to minimize the weighted least-squares objective (\ref{r_objective}).
    \begin{align}
    \label{r_objective}
    \sum_{i \in \mathcal{S}_{\text{train}}} (Z_i - p)^2 \left( \frac{Y_i - \hat{\mu}(X_i)}{Z_i - p} - \hat{\tau}(X_i) \right)^2
    \end{align}
    We used 100 boosting iterations for each model, and all other parameters were set at their default values.
    
    \item \textit{Lasso R-learner}.  This is the same as the XGBoost R-learner, except the models $\hat{\mu}$ and $\hat{\tau}$ were fit using Lasso with 5-fold cross-validation to select the tuning parameter.
    
    \item \textit{Constant}.  In this model, we set $\hat{\tau}$ to be the constant function that always predicts the difference in mean outcomes between treated and control units in the training dataset.
\end{enumerate}

\subsection{Results}

\subsubsection{Causal stacking vs. oracle model selection}

\begin{figure*}
    \centering
    \includegraphics[width=13cm, height=6.5cm]{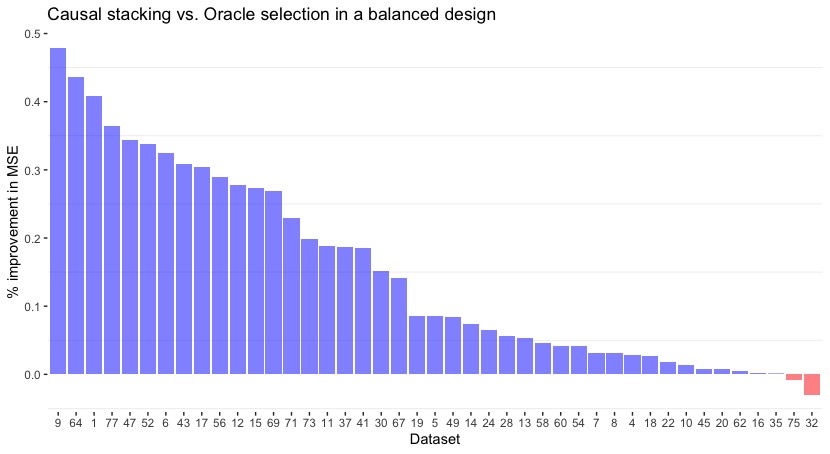}
    \caption{Causal stacking vs. oracle model selection in the balanced design $(p = 0.5)$.  The \% improvement in test-set MSE is defined as $1 - \mathsf{MSE}_{\text{stacking}} / \mathsf{MSE}_{\text{oracle}}$.  In datasets 32 and 75, the decline in MSE is less than 3\%.}
    \label{fig:stacking_vs_oracle}
\end{figure*}

Our simulations suggested three main takeaways in the comparison between causal stacking and oracle model selection:

\textit{1.  Causal stacking outperforms overall}.\\
In every setting of the treatment assignment probability $(p = 0.1, 0.3, 0.5$), causal stacking outperformed oracle model selection on the vast majority of datasets in terms of average mean-squared error on the test set.  Table \ref{tab:stacking_vs_oracle} summarizes these results.

\begin{table}[H]
\centering
\caption{Causal stacking vs. oracle:  \% of datasets won}
\label{tab:stacking_vs_oracle}
\begin{small}
\begin{sc}
\begin{tabular}{ccc}
  \hline
$\P(Z_i = 1)$ & Causal Stacking & Oracle \\ 
  \hline
0.1 &  81\% & 19\% \\ 
0.3&  91\% & 9\% \\ 
0.5 &  95\% & 5\% \\ 
   \hline
\end{tabular}
\end{sc}
\end{small}
\end{table}
    
\textit{2.  Causal stacking shines in balanced experiments}.\\
Figure \ref{fig:stacking_vs_oracle} shows the percent improvement of causal stacking relative to oracle model selection on all 43 datasets under the balanced design ($p = 0.5$).  Causal stacking improves the MSE by as much as 48\%, and never degrades MSE by more than 3\%. 
    
\begin{figure}
    \centering 
    \includegraphics[width = 9cm]{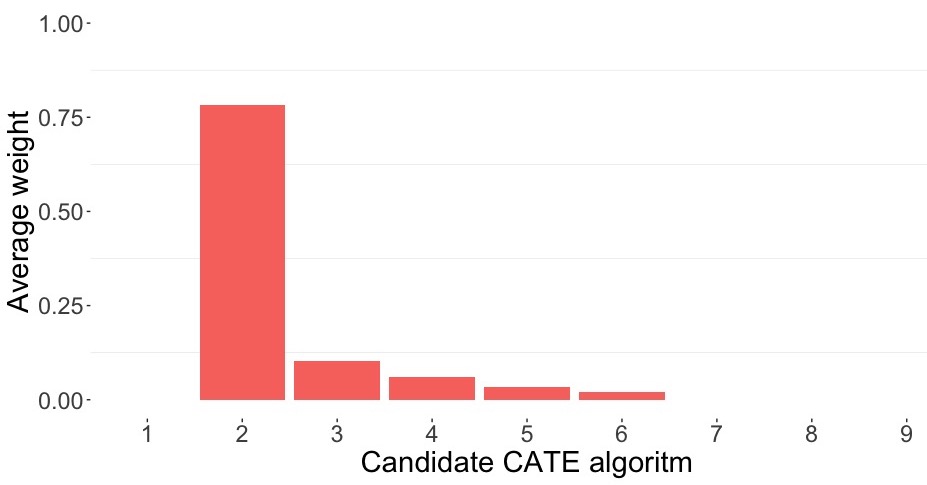}
    \caption{The average causal stacking weights on dataset 5.  Models 1, 7, 8, and 9 received essentially zero weight, due to the sparsity properties of $\ell_1$-constraints.}
    \label{fig:data5_xgboost}
\end{figure}

\textit{3.  No harm from superstars}.\\
In our experiments, we found that a single CATE algorithm (Algorithm 2:  XGBoost T-learner) far outperformed its competition and was chosen by the oracle more often than all other algorithms combined.  For example, the model selection oracle picked the XGBoost T-learner on over 95\% of replications on dataset 5.  Although this situation is often considered to be unfavorable for ensemble methods, we did not find this to be the case.  Figure \ref{fig:data5_xgboost} plots the average weights chosen by causal stacking on dataset 5 for each of the nine candidate CATE algorithms.  Although the dominant model gets the bulk of the weight, contributions from algorithms 3 -- 6 helped causal stacking outperform the model selection oracle on this dataset.  We also performed all of our simulations without the dominant XGBoost T-learner model, and found results essentially identical to those from Table \ref{tab:stacking_vs_oracle}.

\subsubsection{Causal stacking vs. R-Stacking}
\label{CSvsR}
Whether or not causal stacking performs better or R-stacking performs better is more context dependent.  Causal stacking outperformed R-stacking in many simulations, and R-stacking outperformed causal stacking in many others.  We have a few general pieces of advice on how to choose between these two methods:

\begin{table}[H]
\centering
\caption{Causal stacking vs. R-Stacking:  \% of datasets won}
\label{tab:stacking_vs_rlearner}
\begin{small}
\begin{sc}
\begin{tabular}{ccc}
  \hline
$\P(Z_i = 1)$ & Causal Stacking &R-Stacking \\ 
  \hline
0.10 & 35\% & 65\% \\ 
  0.30 & 60\% & 40\% \\ 
  0.50 & 60\% & 40\% \\ 
   \hline
\end{tabular}
\end{sc}
\end{small}
\end{table}

\textit{1. Balanced designs favor causal stacking}.\\
Table \ref{tab:stacking_vs_rlearner} shows the relative win rates of causal stacking vs. R-stacking across the three experimental designs. Causal stacking does better in balanced experiments whereas R-stacking does better in imbalanced experiments.  This agrees with our intuition, since in balanced experiments, R-stacking is (essentially) a special case of causal stacking with regression models $\hat{\mu}_1$ and $\hat{\mu}_0$ constrained to be the same\footnote{The R-stacking procedure uses conic combinations instead of convex combinations.}.  When there are enough samples in both the treatment and control group to estimate different response functions, the added flexibility can reduce bias. In imbalanced experiments, one of the two regression functions is fitted using a relatively small sample, and the added variance may harm performance. In addition, the weights $1/p$ are larger if $p$ is small, so the loss function has higher variance, which degrades the performance of causal stacking.

\begin{figure}[H]
    \centering
    \includegraphics[width=8cm]{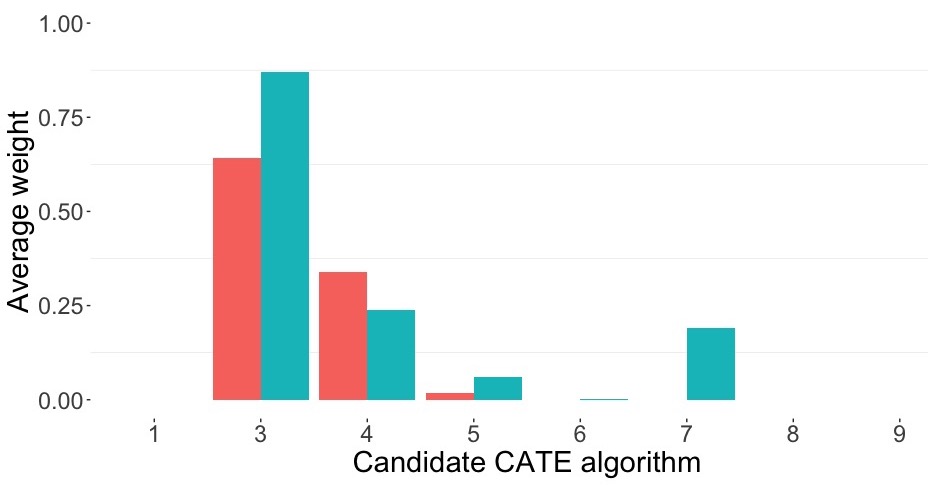}
    \caption{Causal stacking vs. R-stacking weights (red) on dataset 5, without algorithm 2.  The R-stacking model takes advantage of its extra capacity and chooses weights with total sum substantially exceeding one.}
    \label{fig:data5_noxgboost}
\end{figure}

\textit{2.  With worse-performing candidate models, R-stacking is preferred}.\\
In our experiments without the dominant XGBoost T-learner model, the R-stacking procedure performed better than causal stacking (in terms of percent of datasets won) even in balanced experiments. We speculate that this may be due to the capacity control induced by the additional constraint in causal stacking that weights sum to one. With a weaker library of models, greater flexibility may be needed to fit the underlying CATE function.  Figure \ref{fig:data5_noxgboost} shows the average weights of both causal stacking and R-stacking on dataset 5 without Algorithm 2;  a noticeable feature is that the R-stacking procedure uses weights with sum far exceeding one. This extra capacity also allows the R-stacking procedure to employ five candidate models, whereas causal stacking only uses three. In fact, as our next set of experiments show, removing the constraint generally improves the performance in more balanced designs.

\subsubsection{Causal stacking vs. Other variants}
One may consider some other variants of Algorithm~\ref{causal_stacking}. An immediate idea is to remove either one of the constraints or all constraints in \eqref{plug_in}. In our experiments, we found that removing the $\ell_1$ constraint actually improves the performance in more balanced designs. Table \ref{tab:stacking_vs_stacking removing sum to 1} summarizes the results. We see that in more balanced experiments removing the constraint has a significant improvement. Also, removing both the $\ell_1$ constraint and the nonnegative constraint is unstable in our experiments due to co-linearity caused by similarities between some of the models. However, we do see improvements in datasets that do not have the numerical issue. 

\begin{table}[H]
\centering
\caption{Causal stacking vs. Causal Stacking (no $\ell_1$ constraint):  \% of datasets won}
\label{tab:stacking_vs_stacking removing sum to 1}
\begin{small}
\begin{sc}
\begin{tabular}{ccc}
  \hline
$\P(Z_i = 1)$ & Causal Stacking &no $\ell_1$ constraint \\ 
  \hline
0.10 & 58\% & 42\% \\ 
  0.30 & 37\% & 63\% \\ 
  0.50 & 26\% & 74\% \\ 
   \hline
\end{tabular}
\end{sc}
\end{small}
\end{table}

\begin{table}[H]
\centering
\caption{Causal Stacking (no $\ell_1$ constraint) vs. R-Stacking:  \% of datasets won}
\label{tab:stacking removing sum to 1 vs r stacking}
\begin{small}
\begin{sc}
\begin{tabular}{ccc}
  \hline
$\P(Z_i = 1)$ & no $\ell_1$ constraint & R-stacking \\ 
  \hline
0.10 & 35\% & 65\% \\ 
  0.30 & 77\% & 23\% \\ 
  0.50 & 84\% & 16\% \\ 
   \hline
\end{tabular}
\end{sc}
\end{small}
\end{table}

Table \ref{tab:stacking removing sum to 1 vs r stacking} shows the comparison between causal stacking without $\ell_1$ constraint with R-stacking. We see that removing $\ell_1$ constraint makes causal stacking beat R-stacking more often in more balanced designs and still suffer from added variance of fitting two regression functions as explained in the comparison between causal stacking and R-stacking.

\subsubsection{Elaboration on recommendations} Finally, we elaborate on recommendations we gave previously. First, we consider dropping those models that are seldom selected in causal stacking, i.e. we only consider models that are selected quite often. In our experiments, XGBoost T-learner, Random Forest T-learner and Regression tree S-learner are three dominant models. We performed causal stacking with only these three models and compared it with the causal stacking with all nine models. Table \ref{tab:stacking vs stacking few models} shows the results. It is clear from the table that though the other six models were selected only a few times, they do play a role and make the performance better in every setting of the treatment assignment probability we consider.

\begin{table}[H]
\centering
\caption{Causal Stacking vs. Causal Stacking (only model 2, 3 and 4):  \% of datasets won}
\label{tab:stacking vs stacking few models}
\begin{small}
\begin{sc}
\begin{tabular}{ccc}
  \hline
$\P(Z_i = 1)$ & All Models & Only Model 2, 3, 4 \\ 
  \hline
0.10 & 72\% & 28\% \\ 
  0.30 & 65\% & 35\% \\ 
  0.50 & 63\% & 37\% \\ 
   \hline
\end{tabular}
\end{sc}
\end{small}
\end{table}

Second, the quality of two regression models used in \eqref{aipw} matters. We compared causal stacking that uses XGBoost to predict $\{\mu_t\}_{t = 0, 1}$ with causal stacking that uses linear model to predict $\{\mu_t\}_{t = 0, 1}$. In our experiments, we found that the latter one significantly degrades the performance. Table \ref{tab:stacking vs stacking bad mu} summarizes the results.

\begin{table}[H]
\centering
\caption{Causal Stacking (XGBoost $\hat{\mu}_t$) vs. Causal Stacking (linear $\hat{\mu}_t$):  \% of datasets won}
\label{tab:stacking vs stacking bad mu}
\begin{small}
\begin{sc}
\begin{tabular}{ccc}
  \hline
$\P(Z_i = 1)$ & XGBoost $\hat{\mu}_t$ & Linear $\hat{\mu}_t$ \\ 
  \hline
0.10 & 81\% & 19\% \\ 
  0.30 & 93\% & 7\% \\ 
  0.50 & 95\% & 5\% \\ 
   \hline
\end{tabular}
\end{sc}
\end{small}
\end{table}

\section{Discussion}
In \citet{bellkor2007}, the authors of the winning solution to the Netflix prize wrote:
\begin{quote}
\textit{Predictive accuracy is substantially improved when blending multiple predictors.  Our experience is that most efforts should be concentrated in deriving substantially different approaches, rather than refining a single technique.}
\end{quote}
In our paper, we argued that the same wisdom can be applied to the problem of estimating CATE functions. Specifically, we have the following practical recommendations for data analysts working on this problem:
\begin{itemize}
    \item When estimating CATE functions, it is better to fit a diverse array of candidate models and then use model averaging (with either causal stacking or R-stacking) to \textit{combine} these models rather than to use a validation set to \textit{select} a model.
    
    \item On the 43 datasets we studied, X-learner and T-learner were top performers. We recommend that data analysts include these algorithms in their list of candidates.
\end{itemize}
For future work, we would like to better understand which situations lead to causal stacking performing better than R-learner stacking and vice versa.  Although we provided some speculation as to why the R-learner stacking procedure outperformed causal stacking when the dominant XGBoost T-learner model was removed from consideration, we do not yet have any rigorous understanding of this phenomenon.  Other factors to explore include the role of sample size and the number of candidate models $K$;  when the sample size is small or the library of candidate models $K$ is large, we suspect that the extra capacity control in causal stacking will help to control overfitting the validation set. Another interesting direction for future work is to explore more sophisticated ensembling techniques, including causal variants of boosting and bagging. 

\bibliographystyle{plainnat}
\bibliography{references}

\appendix

\section{Proofs}

\subsection{Proof of Proposition 1}
\begin{proof}
We will show the stronger result that the inequality holds even conditional on on the training set $\mathcal{S}_{\text{train}}$. Therefore, we may consider $\hat{\tau}_1, \cdots, \hat{\tau}_K, \hat{\mu}_1$ and $\hat{\mu}_0$ to be fixed functions, and all expectations in this proof will only be taken with respect to the randomness in $\mathcal{S}_{\text{avg}}$.  Define the function $\gamma : \R^K \rightarrow \R^{K + 1}$ by $\gamma(w) = (1, -w)$.  The stacking loss function $\hat{\ell}(w)$ can be written as:
\begin{align*}
\hat{\ell}(w) &= \frac{1}{| \mathcal{S}_{\text{avg}}|} \sum_{i \in \mathcal{S}_{\text{avg}}} ( \hat{\tau}_i - w^{\top} \hat{\tau}_{1:K}(X_i))^2 = \gamma(w)^{\top} \underbrace{\left[ \frac{1}{| \mathcal{S}_{\text{avg}}|} \sum_{i \in \mathcal{S}_{\text{avg}}} \hat{\tau}_{0:K}(X_i) \hat{\tau}_{0:K}(X_i)^{\top} \right]}_{:= \hat{\mathbf{\Sigma}}} \gamma(w)
\end{align*}
where $\hat{\tau}_{0:K}(X_i) = ( \hat{\tau}_i, \hat{\tau}_1(X_i), \cdots, \hat{\tau}_K(X_i))$.  Let $\mathbf{\Sigma} := \E[ \hat{\mathbf{\Sigma}}]$, and notice that the entries of $\hat{\mathbf{\Sigma}}$ are all sample averages of terms bounded in absolute value by $L^2$.  Therefore, Hoeffding's inequality and a union bound imply that:
\begin{align}
\label{entrywise_bound}
\P \left( \max_{1 \leq i, j \leq K + 1} | \hat{\mathbf{\Sigma}}_{ij} - \mathbf{\Sigma}_{ij}| > \sqrt{2} L^2 \sqrt{\frac{\log ( [K + 1]^2 / \delta)}{| \mathcal{S}_{\text{avg}}|}}\right) \leq \delta
\end{align}
On the complement of the event (\ref{entrywise_bound}), the stacking loss $\hat{\ell}(w)$ is close to its expectation $\ell(w) := \E[ \hat{\ell}(w)]$ simultaneously over the set $\Delta_K := \{ w \in \R^K \, : \, w \succeq 0, || w ||_1 = 1 \}$.
\begin{align*} 
\sup_{w \in \Delta_K} | \hat{\ell}(w) - \ell(w) | &= \sup_{w \in \Delta_K} |\gamma(w)^{\top} ( \hat{\mathbf{\Sigma}} - \mathbf{\Sigma}) \gamma(w)|\\
&\leq \left( \sup_{w \in \Delta_K} || \gamma(w) ||_1^2 \right) \left( \max_{1 \leq i, j \leq K + 1} | \hat{\mathbf{\Sigma}} - \mathbf{\Sigma}| \right)\\
&\leq 4  \sqrt{2} L^2 \sqrt{\frac{\log ( [K + 1]^2 / \delta)}{| \mathcal{S}_{\text{avg}}|}}
\end{align*}
This implies that $\ell( \hat{w})$ is not much larger than $\ell(w_{\mathsf{opt}})$, where $w_{\mathsf{opt}} := \displaystyle \argmin_{w \in \Delta_K} \ell(w)$.
\begin{align}
\ell( \hat{w}) &\leq \hat{\ell}( \hat{w}) + | \ell(\hat{w}) - \hat{\ell}( \hat{w})| \notag\\
&\leq \hat{\ell}(w_{\mathsf{opt}}) + | \ell( \hat{w}) - \hat{\ell}( \hat{w})| \notag\\
&\leq \ell(w_{\mathsf{opt}}) + | \ell(w_{\mathsf{opt}}) - \hat{\ell}(w_{\mathsf{opt}})| + | \ell( \hat{w}) - \hat{\ell}( \hat{w})| \notag\\
&\leq \ell(w_{\mathsf{opt}}) + 8 \sqrt{2} L^2 \sqrt{\frac{\log ([K + 1]^2 / \delta)}{| \mathcal{S}_{\text{avg}}|}}
\label{loss_inequality}
\end{align}
To finish the proof, we will use the fact that $\ell( w) = || w^{\top} \hat{\tau}_{1:K} - \tau ||_2^2 + \Var( \hat{\tau}_i)$ for any $w$.  We can see this by first decomposing $\ell(w)$ into three terms:
\begin{align*}
\ell(w) &= \E \left[ \frac{1}{| \mathcal{S}_{\text{avg}}|} \sum_{i \in \mathcal{S}_{\text{avg}}} ( \hat{\tau}_i - w^{\top} \hat{\tau}_{1:K}(X_i))^2 \right]\\
&= \E[ \E \left[ ( \hat{\tau}_i - w^{\top} \hat{\tau}_{1:K}(X_i))^2 | X_i \right]]\\
&= \E[ \E[ (\hat{\tau}_i - \tau(X_i))^2 + (\tau(X_i) - w^{\top} \hat{\tau}_{1:K}(X_i))^2 - 2 ( \hat{\tau}_i - \tau(X_i))(\tau(X_i) - w^{\top} \hat{\tau}_{1:K}(X_i)) | X_i ]]\\
&= \underbrace{\E[ \E[ (\hat{\tau}_i - \tau(X_i))^2 | X_i]]}_{(i)} + \underbrace{|| \tau - w^{\top} \hat{\tau}_{1:K} ||^2_2}_{(ii)} - \underbrace{2 \E[( \tau(X_i) - w^{\top} \hat{\tau}_{1:K}(X_i)) \E[( \hat{\tau}_i - \tau(X_i))|X_i]]}_{(iii)}
\end{align*}
Conditional on $X_i$, $\hat{\tau}_i$ is unbiased for $\tau(X_i$):
\begin{align*}
\E[ \hat{\tau}_i | X_i] &= [\hat{\mu}_1(X_i) - \hat{\mu}_0(X_i)] + \E \left[ \frac{(Y_i - \hat{\mu}_1(X_i)) Z_i}{p} \bigg| X_i \right] - \E \left[ \frac{(Y_i - \hat{\mu}_0)(1 - Z_i)}{1 - p} \bigg| X_i \right]\\
&= [\hat{\mu}_1(X_i) - \hat{\mu}_0(X_i)] + \E[Y_{1i} - \hat{\mu}_1(X_i)| X_i] \E \left[ \frac{Z_i}{p}  \right] - \E \E[ Y_{0i} - \hat{\mu}_0(X_i) | X_i] \E\left[ \frac{1 - Z_i}{1 - p} \bigg| X_i \right]\\
&= \E[ Y_{1i} - Y_{0i} | X_i]\\
&= \tau(X_i)
\end{align*}
Therefore, the term (iii) vanishes and the term (i) simplifies to $\Var( \hat{\tau}_i)$ by the law of total variance.  This establishes the identity $\ell(w) = || w^{\top} \hat{\tau}_{1:K} - \tau ||_2^2 + \Var( \hat{\tau}_i)$, which can be applied on both sides of (\ref{loss_inequality}) to obtain the final result.
\end{proof}

\end{document}